\documentclass[12pt]{article}
\usepackage{amsmath,amssymb,amsfonts}
\usepackage[english]{babel}

\let\p\partial

\def\phi{\varphi}
\let\bsy\boldsymbol
\let\ge\geqslant
\let\le\leqslant
 
\let\t\tilde

\let\ds\displaystyle
\def\n{\nonumber}

\def\be{\begin{equation}}
\def\ee{\end{equation}}
\def\ba{\begin{aligned}} 
\def\ea{\end{aligned}}

\newcounter{theo}
\newcommand{\theo}{\addtocounter{theo}{1}\textbf{Theorem \thetheo.} }
\newcounter{lem}
\newcommand{\lemm}{\addtocounter{lem}{1}\textbf{Lemma \thelem.} }
\newcounter{prop}
\newcounter{rem}
\newcommand{\rem}{\addtocounter{rem}{1}\textbf{Remark \therem.} }
\newcounter{defi}

\newcounter{examp}
\newcommand{\examp}{\addtocounter{examp}{1}\textbf{Example\,\theexamp.} }
\newcounter{cor}

\numberwithin{equation}{section}

\begin{document}
\baselineskip=7mm

\vspace{3mm}

\thispagestyle{empty}
\begin{center}
{\Large\bf Classification of integrable vector  \\ equations of geometric type} \\[2mm]
{\bf A.G. Meshkov$^{a}$,   V.V. Sokolov$^{b,c}$}
\end{center}
\begin{center}
\begin{minipage}[c]{80mm}
{\small
$a).$ Orel State University,  95, Komsomolskaja \\  str., 302026, Orel, Russia
 \\[2mm]
$b).$ Landau Institute for Theoretical Physics, 142432,  Chernogolovka,   Russia 
\\[2mm]
$c).$ Universidade Federal do ABC, 09210-580, Sao Paulo, Brazil  \\ 
}
\end{minipage}

\vspace{1cm}
\begin{minipage}[c]{150mm}
ABSTRACT.  A complete classification of isotropic vector equations of the geometric type that possess higher symmetries is proposed.  New examples of integrable multi-component systems of the geometric type are found. 

\end{minipage}
\end{center}

\section{Introduction}

Consider evolution systems of the form 
\begin{equation}
u^{i}_{t}=u^i_{xxx}+3A^{i}_{jk}({\bsy u})\,u^{j}_{x}\,u^{k}_{xx}+B^{i}_{jkl}({\bsy u})\,u^{j}_{x}\, u^{k}_{x}u^{l}_{x}, \qquad i,j,k,s=1,\dots,N, \label{geom1} 
\end{equation}
where  $\bsy u = (u^1,\dots,u^N)$. 
Here and below, we assume that the summation is carried out over repeated indexes. 

Integrable systems of this type are connected with various geometric and algebraic structures and are of interest by themselves. In addition, the most interesting of them play the role of infinitesimal symmetries for physically important hyperbolic systems of the form
\begin{equation}\label{chiral}
u^i_{xy} = C^i_{jk}(\bsy u) \, u^j_x u^k_y. 
\end{equation}
Having an efficient description of integrable systems  \eqref{geom1}, we can construct a class of integrable systems of the form 
\eqref{chiral} following the approach from the papers \cite{meshsokHyp, MeshSokHy}.

An example of such type integrable 
system provides the following equation \cite{SviSok94}
$$ {\bf U}_t = {\bf U}_{xxx} -
{3\over2}\,{\bf U}_x {\bf U}^{-1} {\bf U}_{xx}-{3\over2}\, {\bf U}_{xx} {\bf U}^{-1} {\bf U}_x
+{3\over2}\, {\bf U}_{x} {\bf U}^{-1} {\bf U}_{x} {\bf U}^{-1} {\bf U}_{x},
$$
where ${\bf U}(x,t)$ is an $m\times m$ matrix. In this case $N=m^2.$ For any $m$ this system has infinitely many local symmetries and conservation laws. 

It is convenient to
rewrite \eqref{geom1} in the following way
\begin{align}
u^i_t&=u^i_3+3 A^i_{jk}(\bsy u)u^j_{x}u^k_{xx}+\left(\frac{\p A^i_{jk} }{\p u^l}+2 A^i_{ls} A^s_{jk}- A^i_{sl} A^s_{jk}+ \beta^i_{jkl} \right) u^j_x u^k_x u^l_x. \label{geeq}
\end{align}

The class of systems \eqref{geeq} is invariant under the arbitrary point transformations ${\bsy u} \rightarrow  {\bf \Phi}({\bsy u})$. 
It is  easy to see that under such a change of coordinates, the functions $A^{i}_{jk}$ and $\beta^{i}_{jkm}$ are transformed just as
components of an affine connection $\Gamma$ and of a tensor $\beta$, respectively. 

\examp 
In the case $N=1$ equation \eqref{geeq} has the form
$$
u_t=u_{xxx} + 3 A(u)\, u_x  u_{xx}  + \Big(A'(u)+A(u)^2+\beta(u)\Big)\, u_x^3. 
$$
Using the symmetry approach (see \cite{MikShaSok91}), one can verify that this equation possesses higher symmetries iff $\beta'=2 A \beta.$ 
By a proper point transformation  of the form $ u \rightarrow \Phi(u)$ the function $A$ can be reduced to zero (for $N=1$ any affine connection is flat) 
and the function $\beta$ becomes a constant. The equation $u_t=u_{xxx}+ {\rm const} \,u_x^3$ is known to be integrable and it is related to the mKdV equation by a potentiation.

Without loss of generality we assume that the tensor $\beta$ is symmetric:
$$
\beta(X,Y,Z)=\beta(Y,X,Z)=\beta(X,Z,Y)
$$
for any vectors  $X,Y,Z$. The functions $\beta^{i}_{jkm}$  are defined by the values of $\beta(X,X,X).$

Suppose a system of the form  \eqref{geeq} has higher symmetries and/or non-degenerate conservation laws and $A^{i}_{jk}=A^{i}_{kj}$, 
i.e. the torsion tensor $T$ is equal to zero. Then\footnote{It was discovered by S.~Svinilupov and V.~Sokolov and was published without proof in the survey
 \cite{habsokyam} dedicated to Sergey Svinolupov.} the corresponding affine connected space is symmetric \cite{Rash} which means that
\begin{equation}\label{Tcon3}
\nabla_X\Big(R(Y,Z,V)\Big)=0,
\end{equation}
where $R$ is the curvature tensor\footnote{We use here the following formula for the curvature tensor:
\begin{align*}\label{}
R^m_{ijk}=\frac{\p}{\p u_j}A^m_{ki}-\frac{\p}{\p u_k}A^m_{ji}+A^m_{js}A^s_{k1}-A^m_{ks}A^s_{ji}.
\end{align*}}. 
Let $$\sigma(X,Y,Z)\stackrel{def}{=}\beta(X,Y,Z)-\frac{1}{3}\Big(R(X,Y,Z)+R(Z,Y,X)\Big).$$ 
Then 
\begin{equation} \label{Tcon0}
\sigma(X,Y,Z)=\sigma(Z,Y,X), 
\end{equation}
\begin{equation}\label{Tcon1}
\nabla_X\Big(\sigma(Y,Z,V)\Big)=0,
\end{equation}
\begin{equation}\label{Tcon2}
R(X,Y,Z)=\sigma(X,Z,Y)-\sigma(X,Y,Z),
\end{equation}
and 
\begin{equation}\begin{array}{l} \sigma(X,\sigma(Y,Z,V),W)-\sigma(W,V,\sigma(X,Y,Z))   
 +\sigma(Z,Y,\sigma(X,V,W))-\sigma(X,V,\sigma(Z,Y,W))=0. \label{jord2}\end{array}
 \end{equation}
The   identities \eqref{Tcon0} and \eqref{jord2} mean that at any point $u$ the tensor $\sigma(u)$ defines a triple Jordan system \cite{Neher,Zel}.

{\bf Conjecture 1.} If a symmetric ($T=0$) affine connection and a tensor $\sigma$ satisfy identities \eqref{Tcon3} -- \eqref{Tcon2} and  \eqref{jord2} then the corresponding\footnote{It is clear that $\beta(X,X,X)=\sigma(X,X,X)$.}  system  
\eqref{geeq} possesses infinitely many local symmetries and conservation laws.

Several integrable systems of the form  \eqref{geeq} that correspond to symmetric connections can be found in 
\cite{SviSok96} but no integrable models corresponding to the case $T\ne 0$ are known. In this paper we construct examples of integrable models  \eqref{geeq} such that $T\ne 0$  and $R=0.$

Our goal is to find all non-triangular integrable systems of the form \eqref{geom1}, which belong to a special class of vector isotropic equations of the form
\begin{equation}\label{evec}
  {\bsy u}_{t} =  {\bsy u}_{xxx}+f_2  {\bsy u}_{xx}+f_1  {\bsy u}_x+f_0  {\bsy u},
\end{equation}
where $\bsy u(x,t)$ is an $N$-dimensional vector and the coefficients $f_i$ are supposed to be  functions of the following six  independent scalar products:
 \begin{equation}\label{dynvar}
(\bsy u,\, \bsy u),\quad (\bsy u, \bsy u_x),\quad  (\bsy u_x, \bsy u_x),\quad  (\bsy u, \bsy u_{xx}),\quad  (\bsy u_x, \bsy u_{xx}),\quad (\bsy u_{xx}, \bsy u_{xx}).
\end{equation}
Equations \eqref{evec} are invariant with respect to the orthogonal group $O_N.$

It is clear that any equation \eqref{evec} whose component form belong to the class of equations \eqref{geom1} has the following structure:
\begin{equation}
\begin{aligned}\label{E1}
\bsy u_t&=\bsy u_{xxx}+a_1u_{[0,1]}\,\bsy u_{xx}+(a_2u_{[0,2]}+a_3u_{[1,1]}+a_4u_{[0,1]}^2)\,\bsy u_x\\[2mm]
&\quad+(a_5u_{[1,2]}+a_6u_{[0,2]}u_{[0,1]}+a_7u_{[1,1]}u_{[0,1]}+a_8u_{[0,1]}^3 )\, \bsy u,
\end{aligned}
\end{equation}
where  
\begin{equation}\label{uvari}
u_{[i,j]}=( \p_x^i\bsy u,\, \p_x^j\bsy u ), \qquad 0\le i\le j 
\end{equation}
and the coefficients $a_i$ are functions in one variable:  $a_i=a_i(u_{[0,0]})$. In this case the components, the torsion and the curvature tensors for the corresponding affine connection are given by 
$$
A(X,Y)=\frac{1}{3}\Big( a_1 (\bsy u,X) Y + a_2 (\bsy u,Y) X + \Big(a_5 (X,Y)+a_6 (\bsy u,X) (\bsy u,Y)\Big) \bsy u\Big),
$$
$$ T(X,Y)=\frac13(a_1-a_2)\Big((\bsy u,X) Y - (\bsy u,Y) X \Big)
$$
and 
$$
\begin{array}{c}
\ds R(X,Y,Z) = \frac{1}{9}\Big(q \,(\bsy u,X) (\bsy u,Z) + p\, (X,Z)  \Big) Y - 
\frac{1}{9}\Big(q \,(\bsy u,X) (\bsy u,Y) + p\, (X,Y)  \Big) Z +  \\[4mm] 
\ds \frac{r}{9}\Big( (\bsy u,Y) (X, Z) -(\bsy u,Z) (X, Y)  \Big) \bsy u, 
\end{array}
$$
where 
$$
p= a_2 a_5\bsy u^2-3 a_2+ 3 a_5, \qquad q=a_2a_6\bsy u^2+a_2^2+3a_6-6a_2',\qquad  r=a_5a_6\bsy u^2+a_5^2-3a_6+6a_5'.
$$

To find all integrable equations \eqref{E1}, we use a version of the symmetry approach developed in  \cite{meshsok} for vector equations.

In Section 2 we discuss necessary conditions \cite{meshsok} of the existence of higher symmetries for vector equations of the form \eqref{evec}.  In Section 3 we present lists of integrable equations \eqref{E1}, formulate and prove classification statements. 
For some of these equations written in components of the vector $\bsy u$ the torsion $T$ is not zero. 
To justify the real integrability of equations found in Section 3,  we detect (see Section 4) auto-B\"acklund transformations for these equations. Each of them is a new integrable semi-discrete model.

\bigskip

{\bf Acknowledgments.} The authors are grateful to E. Ferapontov and P. Leal da Silva for useful discussions. 
 VS was supported by the  state assignment 
No 0033-2019-0006. He is thankful to IHES for its support and hospitality.

\section{Integrability conditions for vector equations}\label{Sec2}
 
 It was shown in \cite{meshsok} that if an equation of the form \eqref{evec} has infinitely many vector higher symmetries
\begin{equation} \label{sym}
{\bsy u}_{\tau} = f_n \bsy u_n+f_{n-1}  \bsy u_{n-1}+\cdots+f_0 \bsy u,\qquad {\rm where} \quad \bsy u_k =\frac{\p^k \bsy u}{\p x^k},
\end{equation}
then an infinite series of special local\footnote{A conservation law $D_t(\rho) = D_x(\theta)$ is called {\bf local} if $\rho$ and $\theta$ are functions of variables \eqref{uvari}.} conservation laws  exists for equation \eqref{evec}. Their  densities $\rho_n,\ n=0,1,\dots$ are called {\bf canonical}. 

The first two canonical densities are given by
\begin{align} \label{ro0}
\rho_0 &= -\frac{1}{3}\,f_2, \\
\rho_1 &=\frac{1}{9}\,f_2^2-\frac{1}{3}\,f_1+\frac{1}{3}\,D_x\,f_2. \label{ro1} 
\end{align}
Using a technique developed in the papers \cite{cheleeliu,M-I}, one can obtain the following recursion formula for other canonical densities for equations of the form (\ref{evec}):
 \begin{align}
\rho_{n+2}&=\frac{1}{3}\biggl[\theta_n-f_0\,\delta_{n,0} -2\,f_2\,\rho_{n+1}- f_2\,D_x\rho_{n} - f_1\,\rho_{n}\biggr] \n \\[2mm]
&-\frac{1}{3}\biggl[f_2\,\sum_{s=0}^{n} \rho_{s}\,\rho_{n-s}+\sum_{0\le s+k\le n}\rho_{s}\,\rho_{k}\, \rho_{n-s-k}+3\sum_{s=0}^{n+1}\rho_{s}\, \rho_{n-s+1}\biggr] \nonumber \\[2mm]
&-D_x\biggl[\rho_{n+1}+\frac{1}{2}\sum_{s=0}^{n}\rho_{s}\,\rho_{n-s}+\frac{1}{3}D_x \rho_{n}\biggr],\qquad n\ge 0. \label{rekkur}
\end{align}
Here the symbol $\delta_{i,j}$ denotes the Kronecker delta and the functions $\theta_i$ are fluxes of the canonical conservation laws \begin{equation}\label{Dx}
D_t\rho_n = D_x\theta_n, \qquad  n=0,1,2,\dots
\end{equation}
In this formula $D_x$ and $D_t$ are the total  derivatives of $x$ and $t$, respectively. For brevity, we call relation (\ref{Dx})  $\rho_n$-integrability condition.

Using formulas (\ref{ro0})--(\ref{rekkur}), one can obtain the next density
\begin{align} \label{ro2}
\rho_2 = -\frac{1}{3}\,f_0+\frac{1}{3}\,\theta _0-\frac{2}{81}\,f_2^3+ \frac{1}{9}\,f_1\,f_2-D_x\biggl(\frac{1}{9}\,f_2^2 + \frac{2}{9}\,D_x\,f_2 -\frac{1}{3}\,f_1\biggr)
\end{align}
and so on. Notice that the density $\rho_n,\,n\ge2$  depends on the coefficients of (\ref{evec}) and on the fluxes $\{\theta _0,\theta_1,\dots  \theta_{n-2}\}$. These fluxes are to be calculated from the previous conditions (\ref{Dx}).

To eliminate the function $\theta_n$ from \eqref{Dx} one can apply  the variational derivative
\begin{equation}\label{vars}
\frac{\delta}{\delta \bsy u} = \sum_{0\le i\le j}\left[(-D_x)^i\left(\bsy u_j \frac{\p }{\p u_{[i,j]}}\right)+(-D_x)^j\left(\bsy u_i \frac{\p }{\p u_{[i,j]}}\right)\right] .
\end{equation}
to both sides of (\ref{Dx}) and use the fact that   
$$ 
\frac{\delta (D_x g)}{\delta \bsy u} = 0
$$
for any function $g$ (see, for example \cite{olver}, chapter 4) to obtain
\begin{equation}\label{varderform}
\frac{\delta }{\delta \bsy u}(D_t \rho_n) = 0, \qquad n=1,2,\dots \,\,.
\end{equation}
Conditions (\ref{varderform}) are  most efficient for the cases $n=1,2$ since (\ref{ro0}) and (\ref{ro1}) do not depend on 
$\theta_i.$

\section{Classification of integrable equations (\ref{E1})}

We are searching for non-triangular integrable equations of the  form \eqref{E1}. In this section, {\bf integrability} means the existence of an infinite sequence of higher symmetries \cite{olver, MikShaSok91} of the form (\ref{sym}).

Some equations  \eqref{evec} become triangular in the spherical coordinates, which is defined by the formulas 
$$\bsy u=R\,\bsy v,\qquad |\bsy v|=1, \qquad \mbox{\rm where} \quad R=|\bsy u|.$$ Let 
$$
v_{[i,j]}=( \p_x^i\bsy v,\, \p_x^j\bsy v),\qquad  i\le j. 
$$ 
Since $v_{[0,0]} =1$, we have $D_x(v_{[0,0]})=2v_{[0,1]}=0$. Moreover, $D_x v_{[0,1]}=v_{[0,2]}+v_{[1,1]}=0$, i.e. $v_{[0,2]}=-v_{[1,1]}$ and so on. It is clear that 
all variables $v_{[0,k]}$ can be expressed in terms of the variables $v_{[i,k]},\ 1\le i\le k<\infty$. 

We call equation \eqref{evec} {\bf triangular} if it can be rewritten in the  spherical coordinates as
$$
\begin{array}{l}
  {\bsy v}_{t} =  {\bsy v}_{xxx}+g_2  {\bsy v}_{xx}+g_1  {\bsy v}_x+g_0  {\bsy v}, \\[3mm]
 R_t = R_{xxx} + S(v_{[1,1]},v_{[1,2]},v_{[2,2]}, R, R_x, R_{xx}),
\end{array}
$$
where the coefficients $g_i$ depend on $v_{[1,1]},v_{[1,2]},v_{[2,2]}$ only. 

\subsection{Classification statements}
 
The class of equations of the form (\ref{E1}) is invariant with respect to the point transformations of the form 
\begin{equation}\label{pointgen}
\bsy u=\bsy v\, \phi(v_{[0,0]}).
\end{equation}
Under such a transformation the coefficient $a_1$ changes as follows:
$$
\t a_1(v_{[0, 0]})=2\,\phi^{-1}\phi'\big(v_{[0, 0]}\,a_1\phi^2+3\big)+a_1\,\phi^2, \qquad \mbox{where} \quad a_1(u_{[0,0]})=a_1\big(v_{[0,0]}\phi^2\big).
$$
It easy to see that if  $a_1=-3\,{u_{[0,0]}}^{-1},$ then we obtain $\t a_1=-3\,{v_{[0,0]}}^{-1}$. For any function $a_1$ different from $-3\,{u_{[0,0]}}^{-1}$ we can choose the function  $\phi$ 
such that $\t a_1$ vanishes.
Thus, up to the point transformations we have two non-equivalent cases: 
$$
{\bf 1.} \quad a_1=0 \qquad \mbox{and} \qquad {\bf 2.} \quad  a_1=-\frac{3}{{u_{[0,0]}}}. 
$$

\theo Any non-triangular integrable equation of the form (\ref{E1}) with  $a_1=0$ can be reduced to one of equations from the following List 1 by a scaling  of the form $\bsy u\to \lambda\bsy u$. 

{\bf List 1.}  
\begin{align}
\bsy u_t&=\bsy u_{xxx}+\frac{3\lambda}{z} \, \bsy u_x\left(\frac{u_{[0,1]}^2 }{z+u_{[0,0]}}-u_{[1,1]}\right)+\frac3z\bsy u\,F,\ 
 \qquad  {\rm where} \quad \lambda = 1 \quad {\rm or} \quad \lambda = \frac{1}{2},  \label{e1} \\[3mm] %
%
%
\bsy u_t&=\bsy u_{xxx}-\frac3z\,\bsy u_x\left(\frac{zu_{[0,2]}}{z+u_{[0,0]}}-\frac{u_{[0,0]}u_{[1,1]}}{z+u_{[0,0]}}+\frac{u_{[0,0]}u_{[0,1]}^2}{(z+u_{[0,0]})^2}\right)
+\frac3z\,\bsy u F,  \label{e2} \\[3mm]
\bsy u_t&=\bsy u_{xxx}-\frac{3}{z}\,\bsy u_x\left(\frac{zu_{[0,2]}}{z+u_{[0,0]}}+\frac{(z-u_{[0,0]})u_{[1,1]}}{2\,(z+u_{[0,0]})}-\frac{(2z-u_{[0,0]})u_{[0,1]}^2}{2\,(z+u_{[0,0]})^2}\right)
+\frac3z\,\bsy u F,  \label{e3}  \\
\intertext{where}
&F=u_{[0,1]}\frac{u_{[0,2]}+u_{[1,1]}}{z+u_{[0,0]}}-u_{[1,2]}-\frac{u_{[0,1]}^3}{(z+u_{[0,0]})^2}. \n 
\end{align}
Here  $z \ne 0$ is an arbitrary parameter. 

\theo Any non-triangular integrable equation of the form  (\ref{E1}) with $\ds a_1=-\frac{3}{u_{[0,0]}}$ can be reduced to one of the equations from the following List 2 by a point transformation of the form  \eqref{pointgen}.

{\bf List 2.} 
\begin{align}
\bsy u_t&=\bsy u_{xxx} -3\bsy u_{xx}\frac{u_{[0,1]}}{u_{[0,0]}}-3\bsy u_x\left(\frac{u_{[1,1]}}{u_{[0,0]}}-\frac{u_{[0,1]}^2}{u_{[0,0]}^2}\right),  \label{e4}  \\[3mm]
\bsy u_t&=\bsy u_{xxx}-3\bsy u_{xx}\frac{u_{[0,1]}}{u_{[0,0]}}-\frac32\bsy u_x\left(\frac{u_{[1,1]}}{u_{[0,0]}}-2\frac{u_{[0,1]}^2}{u_{[0,0]}^2}\right),  \label{e5} 
\end{align}
\begin{align}
\bsy u_t&=\bsy u_{xxx}-3\bsy u_{xx}\frac{u_{[0,1]}}{u_{[0,0]}}-\frac{3}{2}\,\bsy u_{x}\left(2\frac{u_{[0,2]}}{u_{[0,0]}}+\frac{u_{[1,1]}}{u_{[0,0]}}\right) 
+3\bsy u\left(\frac{u_{[1,2]}}{u_{[0,0]}}-\frac{u_{[0,1]}u_{[1,1]}}{u_{[0,0]}^2}+\frac43\,\frac{u_{[0,1]}^3}{u_{[0,0]}^3}\right).         \label{e6}    
\end{align}

\rem Equation \eqref{e6} is equivalent to the equation 
\begin{align}\label{e7}
&\bsy u_t=\bsy u_{xxx}-3\bsy u_{xx}\frac{u_{[0,1]}}{u_{[0,0]}} -\frac32\bsy u_{x}\left( 2\frac{u_{[0,2]}}{u_{[0,0]}} +\frac{u_{[1,1]}}{u_{[0,0]}}-4\frac{u_{[0,1]}^2}{u_{[0,0]}^2}\right)  
 +3\bsy u\left(\frac{u_{[1,2]}}{u_{[0,0]}}-\frac{u_{[1,1]}u_{[0,1]}}{u_{[0,0]}^2}\right) 
\end{align}
found in \cite[formula (59)]{habsokyam}.

\rem Using the formulas from Introduction, one can verify that for equations \eqref{e1}  and \eqref{e6} the torsion  $T$ is equal to zero while
for equations \eqref{e2}--\eqref{e5} we have $T\ne 0, \,\,R=0$. 

%
%
%
%
%
\subsection{Proof of Theorem 1}\label{PrT1}

 The equation under consideration is the following:
\begin{equation}
\begin{aligned}\label{E1-0}
\bsy u_t&=\bsy u_{xxx}+(a_2u_{[0,2]}+a_3u_{[1,1]}+a_4u_{[0,1]}^2)\,\bsy u_x + \\
&\quad (a_5u_{[1,2]}+a_6u_{[0,2]}u_{[0,1]}+a_7u_{[1,1]}u_{[0,1]}+a_8u_{[0,1]}^3 )\, \bsy u.
\end{aligned}
\end{equation}
For such equations the canonical densities (\ref{ro0}),  (\ref{ro1}),  and (\ref{ro2})  are given by
\begin{align*}
\rho_0&=0,\qquad \rho_1=-\frac{1}{3}\big(a_2u_{[0,2]}+a_3u_{[1,1]} +a_4u_{[0,1]}^2\big), \\[3mm]
\rho_2&= -\frac{1}{3}\,\big(a_5u_{[1, 2]}+a_6u_{[0, 2]}u_{[0, 1]}+a_7u_{[1, 1]}u_{[0, 1]}+a_8u_{[0, 1]}^3\big)
+\frac{1}{3}\,D_x(a_2u_{[0, 2]}+a_3u_{[1, 1]}+a_4u_{[0, 1]}^2).
\end{align*}

Consider the $\rho_1$-condition. 
 The equality  \eqref{varderform} with $n=1$ has the form $\sum_{i=0}^4 q_i \bsy u_i = 0,$ where  
$$
q_4 = 4\,\,u_{[0, 1]}(a_2'-a_3').
$$
Hence, $a_3=a_2+c_1,$ where $c_1$ is a constant. Eliminating $a_3$,
we find that $q_3$ vanishes, which allows us to write the coefficient $q_2$ as
$$
q_2= (u_{[0, 3]}+3u_{[1, 2]})\big((a_5u_{[0, 0]}-3)(2\,a'_2-a_4)-c_1(a_2+a_5)\big)+ u_{[0, 1]}\big(u_{[0,2]}F_1+u_{[1,1]}F_2+u_{[0, 1]}^2F_3\big),
$$
where $F_i=F_i(u_{[0, 0]})$. 
The functions $F_i$ are too cumbersome to be shown explicitly here while the difference $F_1-F_2$ is very short:
$$
F_1-F_2 = c_1(a_5'+2\,a_6-2\,a_7).
$$
Thus from (\ref{varderform}) with $n=1$ we have obtained three simple relations
\begin{align}\label{eq1}
a_3=a_2+c_1,\quad (a_5u_{[0, 0]}-3)(2\,a'_2-a_4)-c_1(a_2+a_5)=0,\quad c_1(a_5'+2\,a_6-2\,a_7)=0.
\end{align}

Consider now the $\rho_2$-condition.  We  obtain  
$$
\frac{\delta }{\delta \bsy u}(D_t \rho_2)=\sum_{i=0}^4 p_i \bsy u_i = 0,
$$
where 
$$
p_4= \Big((u_{[0, 2]}+u_{[1, 1]})\,(a_5'+a_6-a_7)+2\,u_{[0, 1]}^2(a_5''+a_6''-a_7')\Big). 
$$
Equating $p_4$ to zero, we find $a_7=a_5'+a_6$ and conclude that this implies $p_3=0$. Substituting $a_7$ into third of equations (\ref{eq1}), we obtain that $c_1 a_5' = 0$.   Equating now $p_0$ to zero, 
we find one more simple relation $a'_6-a_8=0$. So the $\rho_2$-condition implies  
\begin{align}\label{eq2}
a_7=a'_5+a_6, \qquad c_1 (a_6-a_7) = 0, \qquad a'_6-a_8 = 0.
\end{align}

Several more useful relations can be derived from the $\rho_4$-condition.  The density $\rho_4$ has the following structure: 
$$
\rho_4=\frac13\theta_2-\frac13 D_x(\theta_1) +R,
$$
where $R$ does not depend on $\theta_1$ and $\theta_2$. The term with $\theta_1$ disappears when we apply the variational derivative   in the formula \eqref{varderform} with $n=4$. So to use the $\rho_4$-condition, we have to specify the form of the function   $\theta_2$ only. 

Using (\ref{eq2}), we obtain  that  $\rho_2$ is trivial:   $\rho_2=D_x(S)$, where  
$$
S=\frac16\, \Big(2a_4u_{[0, 1]}^2-a_6u_{[0, 1]}^2+2a_2u_{[0, 2]}+2a_2u_{[1, 1]}-a_5u_{[1, 1]}+2c_1u_{[1, 1]}\Big).
$$
Therefore, $\theta_2=D_t(S)$. 
Taking into account this expression for $\theta_2,$  we find that 
$$
\frac{\delta }{\delta \bsy u}(D_t \rho_4)=\sum_{i=0}^6 r_i \bsy u_i,
$$
where  
$$
r_6 = 2\,\Big(2\,a_5''u_{[0, 1]}^2+a_5'(u_{[0, 2]}+u_{[1, 1]})\Big).
$$
This means that $a_5=c_2$, where $c_2$ is a constant. Substituting $a_5=c_2$ into  the second equation of (\ref{eq1}), we obtain
\begin{align}
  (c_2 u_{[0, 0]}-3)(2\,a'_2-a_4)+(a_2+c_2)c_1=0,\qquad \text{or} \qquad  a_4 = 2\,a'_2-\frac{c_1(a_2+c_2)}{c_2u_{[0, 0]}-3}. \label{eq2a}
\end{align}

 Using (\ref{eq1}), (\ref{eq2}), (\ref{eq2a}), we express all coefficients in (\ref{E1-0}) in terms of $a_2,a_6, c_1, c_2.$ Then
the coefficient $r_5$ vanishes and $r_4$ turns into  
$$
r_4 =\big(u_{[0, 4]}+4\,u_{[1, 3]}\big)\big(a_6c_2u_{[0, 0]}+c_2^2-3\,a_6\big)+4\, u_{[[0, 1]}u_{[0, 3]}\,c_1\big(a'_2(c_2\,u_{[0, 0]}-3)+a_2c_2\big). 
$$
The equation $r_4=0$ is then equivalent to relations
\begin{equation}\label{eq2b}
a_6 = -\frac{c_2^2}{c_2\,u_{[0, 0]}-3} \qquad \text{ and }\qquad  c_1\big(a'_2(c_2\,u_{[0, 0]}-3)+a_2c_2\big)=0 
\end{equation}
and we have proved the following:

\lemm
Any integrable equation (\ref{E1-0}) has the form
\begin{equation}\label{Eq0}
\begin{aligned}
\bsy u_t&=\bsy u_{xxx}+\bsy u_x\left(a_2u_{[0, 2]}+(a_2+c_1)u_{[1, 1]}+\Big(2\,a'_2-c_1\frac{a_2+c_2}{c_2u_{[0, 0]}-3}\Big)u_{[0, 1]}^2\right) + \\
&\ \ \bsy u\left(c_2u_{[1, 2]}-c_2^2u_{[0, 1]}\frac{u_{[0, 2]}+u_{[1, 1]}}{c_2u_{[0, 0]}-3}+\frac{c_2^3u_{[0, 1]}^3}{(c_2u_{[0, 0]}-3)^2}\right),
\end{aligned}
\end{equation}
where $a_2=a_2(u_{[0, 0]})$ and $c_i$ are constants. 

Let us consider the following two  branches
$$
{\bf A.}\ \ c_1=0\qquad {\rm and} \qquad {\bf B.}\ \ c_1\ne0.
$$

In Case {\bf A} the $\rho_5$-condition leads to $c_2=0$ and $a_2=0$ and the linear equation $\bsy u_t=\bsy u_3$ appears.

 Case {\bf B} we separate into two following subcases:
$$
{\bf B.1.}\ \ a_2=0 \qquad {\rm and} \qquad {\bf B.2.}\ \ a_2\ne0.
$$

Consider Case {\bf B.1.} If $c_2=0$, then we arrive at the equation  $\bsy u_t=\bsy u_3+\bsy u_1\,c_1u_{[1, 1]}$. This equation is not integrable since the $\rho_5$-condition leads to a contradiction.

If  $c_2\ne0$,  then equation (\ref{Eq0}) coincides with equation (\ref{e1}), where 
$$
 c_2=-\frac{3}{z}, \qquad  c_1=-\frac{3\lambda}{z}.
$$ 
The $\rho_5$-condition gives rise to the following equation 
$$(2\lambda -1)(\lambda -1)=0.$$

Consider Case {\bf B.2.} The coefficient $r_0$ in the $\rho_4$-condition is given by
$$
r_0= -\frac{4}{9}\, u_{[0, 1]}u_{[1, 6]}c_1\big(3\,a_2'-a_2^2\big)-\frac{4\,u_{[0, 2]}u_{[0, 6]}}{3\,(c_2u_{[0, 0]}-3)^2}\,a_2c_1\big(a_2(c_2u_{[0, 0]}-3)+3\,c_2\big).
$$
Therefore,
\begin{align*}
c_1\big(3\,a'_2-a^2_2\big)=0,\qquad a_2 c_1\Big(a_2 (c_2u_{[0, 0]}-3)+3\,c_2\Big) = 0
\end{align*}
and we have
$$
a_2 = -\frac{3\,c_2}{c_2u_{[0, 0]}-3}.
$$
Then $\rho_5$-condition provides the following equation 
$$(c_1+c_2)(2\,c_1+c_2)=0.$$  
The two possibilities $c_1=-c_2$ and $\ds c_1=-\frac{c_2}{2}$ correspond to equations (\ref{e2}) and (\ref{e3}), where $\ds c_2=-\frac{3}{z}$. \qquad $\square$

\rem We have verified that all equations of List 1 satisfy the $\rho_{n}$-conditions with $n \le 7 $. It turns out that $\rho_n$, where $n = 0,2,4,6$, are total $x$-derivatives. In accordance with a general statement from \cite{meshsok} this is an indication of the existence of infinite series of local conservation laws. 
The canonical conservation laws, corresponding to $n= 1,3,5,7$, have the orders $1,2,3,4,$ respectively. Moreover, each  equation from List 1 possesses a fifth order symmetry.

\subsection{Proof of Theorem 2}\label{PrT2}

Consider equations of the form
\begin{equation}
\begin{aligned}\label{E1-00}
\bsy u_t&=\bsy u_{xxx}-3\frac{u_{[0,1]}}{u_{[0,0]}}\bsy u_{xx}+(a_2 u_{[0,2]}+a_3u_{[1,1]}+a_4u_{[0,1]}^2)\,\bsy u_x+\\
&\quad(a_5u_{[1,2]}+a_6u_{[0,2]}u_{[0,1]}+a_7u_{[1,1]}u_{[0,1]}+a_8u_{[0,1]}^3 )\, \bsy u.
\end{aligned}
\end{equation}
The simplest canonical densities  (\ref{ro0}),  (\ref{ro1}),  and (\ref{ro2}) are given by
\begin{align*}
\rho_0 &=\frac{1}{2}\,D_x(u_{[0,0]}),\qquad \rho_1= \frac{u_{[0,1]}^2}{u_{[0,0]}^2}-\frac{1}{3}\Big(a_2u_{[0,2]}+a_3u_{[1,1]}+a_4u_{[0,1]}^2\Big)-D_x\Big(\frac{u_{[0, 1]}}{u_{[0, 0]}}\Big), \\[2mm]
\rho_2 &=\frac{1}{3}D_x\left(\frac{u_{[0, 2]}}{u_{[0, 0]}}-\frac{7}{2}\,\frac{u_{[0, 1]}^2}{u_{[0, 0]}^2}-\frac{u_{[1, 1]}}{2\,u_{[0, 0]}}+2\,D_x\Big(\frac{u_{[0, 1]}}{u_{[0, 0]}}\Big)+a_2u_{[0, 2]}+a_3u_{[1, 1]}+a_4u_{[0, 1]}^2\right).
\end{align*}
 
Using the same line of reasoning as in Section \ref{PrT1}, we derive short relations from the   $\rho_1$ -- $\rho_5$ conditions.   Namely, it follows from the $\rho_1$-condition that 
\begin{equation}\label{Q1}
a_3=a_2+c_1 u_{[0, 0]}^{-1},
\end{equation}
\begin{equation}\label{Q2}
 (a_5u_{[0, 0]}-3)(2\,u_{[0, 0]}^2a'_2-a_4u_{[0, 0]}^2+3) = c_1 u_{[0, 0]}(a_2+a_5),
\end{equation} 
where $c_1$ is a constant. 
The  $\rho_3$ and $\rho_4$-conditions implies 
\begin{equation}\label{Q3}
c_1\Big(6\,a_2'u_{[0, 0]}-(a_2u_{[0, 0]}+3)(a_6u_{[0, 0]}+a_5)-a_2(a_2u_{[0, 0]}-3)\Big)=0, 
\end{equation} 
\begin{equation}\label{Q4}
c_1\Big((a_7-a_6)u_{[0, 0]}+a_5\Big)(a_2u_{[0, 0]}+3)=0,
\end{equation}
and the  $\rho_5$-condition leads to the following relations: 
\begin{equation}\label{Q5}
\Big(a_5(a_2u_{[0, 0]}+3)-3\,a_2\Big)(a_2u_{[0, 0]}+3)=0, \\  
\end{equation}
\begin{equation}\label{Q6}
2\,c_1\,u_{[0, 0]}(a_5(a_2u_{[0, 0]}+3)-3\,a_2)=(c_1+3)(2c_1+3),\ \\ 
\end{equation}
\begin{equation}\label{Q7}
2\,u_{[0, 0]}^2(a_2u_{[0, 0]}+3)(a'_5+a_6-a_7) + (a_5u_{[0, 0]}-3)(a_4u_{[0, 0]}^2+2a_2u_{[0, 0]}-3)+c_1u_{[0, 0]}(a_2+a_5)=0.  
\end{equation}
It follows from (\ref{Q6}) that $c_1\ne0$ and we may reduce (\ref{Q3}) and (\ref{Q4}) by the factor $c_1$.  

Let us simplify the equation \eqref{E1-00} by an appropriate point transformation of the form \eqref{pointgen}. It is more convenient for computations to rewrite it as
\begin{equation}\label{point}
\bsy u=\Big(\frac{f}{v_{[0, 0]}}\Big)^{1/2}\bsy v,\qquad u_{[0, 0]}=f(v_{[0, 0]}).
\end{equation}
One can verify that under this transformation the coefficient $a_2$ transforms as
\begin{equation}\label{tildea}
 \tilde a_2 = \frac{\p f}{\p v_{[0, 0]}}\frac{a_2 f + 3}{f} - \frac{3}{v_{[0, 0]}}.
\end{equation}
It follows from this formula that we can reduce $a_2$ to zero with the exception of the case $\ds a_2(u_{[0, 0]}) = - \frac{3}{u_{[0, 0]}}.$
 
 {\bf  Case A:} \,\, $a_2=0.$    It follows from \eqref{Q5}, \eqref{Q3} and \eqref{Q4} that $a_5=a_6=a_7=0$.  From (\ref{Q2}) we obtain $\ds a_4=\frac{3}{u_{[0, 0]}^2}.$ 
Moreover, relation (\ref{Q6}) leads to $(c_1+3)(2c_1+3)=0$.  Substituting $a_i,\ 2\le i\le 7$ into the $\rho_1$-condition, we obtain that $a_8'=-3\,a_8 u_{[0, 0]}$ or $a_8 = k/u_{[0, 0]}^3$.  
Finally,  $\rho_5$-condition gives rise to  $k=0$.

In the case $c_1=-3$ we arrive at equation \eqref{e4} while $\ds c_1=-\frac{3}{2}$ leads to equation \eqref{e5}. 
It follows from \eqref{tildea} that in Case A the only admissible point transformations are ${\bf u} \to {\rm const} \, {\bf u} $ and therefore  equations \eqref{e4} and \eqref{e5} are non-equivalent.

 {\bf Case B:} \,\, $\ds a_2 = -\frac{3}{u_{[0, 0]}}.$ Taking into account \eqref{Q1}, we find that the
equation has the following form
\begin{equation}
\begin{aligned}\label{E1-01}
\bsy u_t&=\bsy u_{xxx}-3\frac{u_{[0,1]}}{u_{[0,0]}}\bsy u_{xx}+\bsy u_x\left(a_4u_{[0, 1]}^2-3\frac{u_{[0, 2]}}{u_{[0, 0]}}+(c_1-3)\frac{u_{[1, 1]}}{u_{[0, 0]}}\right)+\\
&\qquad\bsy u\Big(a_5u_{[1,2]}+a_6u_{[0,2]}u_{[0,1]}+a_7u_{[1,1]}u_{[0,1]}+a_8u_{[0,1]}^3 \Big) .
\end{aligned}
\end{equation}
Relations \eqref{Q2} and \eqref{Q6} can be rewritten as 
\begin{equation}\label{eq9}
(c_1-3)(2\,c_1-3)=0, \qquad   (a_5u_{[0, 0]}-3)(a_4u_{[0, 0]}^2+c_1-9)=0.
\end{equation}
For equations of the form \eqref{E1-01}
the $\rho_1$-condition provides  the following additional relations:
\begin{align}
&(a_4u_{[0, 0]}^2+c_1-9)(a_6u_{[0, 0]}^2-a_7u_{[0, 0]}^2-3)=0, \\[2mm] 
& a_4' =\frac{a_6}{3u_{[0, 0]}}\,(a_4u_{[0, 0]}^2+c_1-9)-2\frac{a_4}{u_{[0, 0]}}.  \label{eq10} \
\end{align}

Moreover it follows from the $\rho_5$-condition that
\begin{align}
&(a_4u_{[0, 0]}^2+c_1-9)(2\,c_1-3)=0,  \\[2mm]
&(a_4u_{[0, 0]}^2+c_1-9)(u_{[0, 0]}^4a_6^2+6\,a_6'u_{[0, 0]}^3-9\,a_8u_{[0, 0]}^3-6\,a_4u_{[0, 0]}^2-6\,a_6u_{[0, 0]}^2+36)=0.  
\label{eq111}
\end{align}

Under transformations \eqref{point} the coefficient $a_4$ changes as follows:
$$
\tilde a_4 = \frac{f'^2}{f^2} \Big( a_4 f^2+c_1 -9\Big) + \frac{9-c_1}{v_{[0, 0]}^2}.
$$
The condition $\t a_4=0$ is a differential equation for $f$, which has a non-constant solution except for the case  $a_4u_{[0, 0]}^2+c_1-9=0$. So we arrive at the following two cases:
$$
 {\bf B.1.} \quad a_4 = 0 \qquad \text{and} \qquad {\bf B.2.} \quad a_4 = \frac{9-c_1}{u_{[0, 0]}^2}.
$$

In the case {\bf Case B.1.} it follows from  (\ref{eq9}) -- (\ref{eq111}) that 
$$
a_4=0,\qquad  a_5=\frac{3}{u_{[0, 0]}},\qquad a_6=0,\qquad a_7=-\frac{3}{u_{[0, 0]}^2},\qquad a_8=\frac{4}{u_{[0, 0]}^3},\qquad c_1=\frac{3}{2}.
$$

Substituting all these coefficients into equation \eqref{E1-01}, we obtain equation \eqref{e6}.

Consider the {\bf Case B.2.} According to equations (\ref{eq9}) and (\ref{eq10}) we have the following equation:
\begin{align}\label{eq13}
\bsy u_t&=\bsy u_3-3\bsy u_2\frac{u_{[0,1]}}{u_{[0,0]}}+\bsy u_1\left((9-c_1)\frac{u_{[0,1]}^2}{u_{[0,0]}^2}-3\frac{u_{[0,2]}}{u_{[0,0]}}+(c_1-3)\frac{u_{[1,1]}}{u_{[0,0]}}\right) \n \\
&\quad+\bsy u\left(a_5u_{[1,2]}+a_6u_{[0,1]}u_{[0,2]}+a_7u_{[0,1]}u_{[1,1]}+a_8u_{[0,1]}^3\right), 
\end{align}
where  $(c_1-3)(2\,c_1-3)=0$.  It can be verified that in the spherical coordinates equation \eqref{eq13} has the form 
\begin{align}
&\bsy v_t =\bsy v_3+ c_1\bsy v_1v_{[1, 1]}+3 \bsy v v_{[1, 2]}, \qquad (c_1-3)(2\,c_1-3)=0, \ \label{eq14} \\[3mm]
&R_t = R_3+\frac{R_1R_2}{R}(R^4b_6+R^2b_5-6)-R_1v_{[1, 1]}(R^4b_6-R^4b_7-R^2b_5-c_1)\n \\[2mm]
 &\quad \qquad \qquad+\frac{R_1^3}{R^2}(R^6b_8+R^4b_7+6)+R v_{[1, 2]}(R^2b_5-3), \label{eq15}
\end{align}
where $b_i(R) = a_i(u_{[0,0]})\equiv a_i(R^2)$.  So, the system \eqref{eq13} is triangular.  \qquad $\square$

\rem Both equations \eqref{eq14} are integrable equations on the sphere \cite{meshsok}.  They have infinitely many conservation laws depending on the variables  $v_{[i,j]}.$  This is a reason why all conditions from Section \ref{Sec2} are satisfied for any functions $a_5-a_8$. However, we can use the geometric integrability conditions \eqref{Tcon3} - \eqref{jord2} for the classification of triangular systems \eqref{eq13} (see Appendix \ref{app2}).

\rem It turns out that the equations of List 1 can be simplified by the point transformation \eqref{point} with 
$$
 f = \frac{z\,v_{[0, 0]}}{a-v_{[0, 0]}}, 
$$
where $a\ne0$ is an arbitrary constant. As a result, the coefficients of $\bsy u$ vanish and the equations \eqref{e1}, \eqref{e2} and 
\eqref{e3} transform to equations 
\begin{align}\label{e1-aa}
\bsy u_t=\bsy u_3-3\bsy u_2\frac{u_{[0, 1]}}{a+u_{[0, 0]}}-3\bsy u_1\left(\frac{u_{[0, 2]}}{a+u_{[0, 0]}}-\frac{u_{[1, 1]}(\lambda -1)}{a+u_{[0, 0]}}+\frac{u_{[0, 1]}^2(\lambda-3)}{(a+u_{[0, 0]})^2}\right), 
\end{align}
 where $ \lambda = 1$ or  $\ds \lambda = \frac{1}{2},$
\begin{align}\label{e2-aa}
\bsy u_t=\bsy u_3-3\bsy u_2\frac{u_{[0, 1]}}{a+u_{[0, 0]}}-3\bsy u_1\left( \frac{u_{[1, 1]}}{a+u_{[0, 0]}}-\frac{u_{[0, 1]}^2}{(a+u_{[0, 0]})^2}\right), 
\end{align}
and 
\begin{align}\label{e3-aa}
\bsy u_t=\bsy u_3-3\bsy u_2\frac{u_{[0, 1]}}{a+u_{[0, 0]}}-\frac32\bsy u_1\left(\frac{u_{[1, 1]}}{a+u_{[0, 0]}}-\frac{2u_{[0, 1]}^2}{(a+u_{[0, 0]})^2}\right),
\end{align}
respectively. Equations written in this form appeared in \cite{M-Bal} (see formulas (3.12), (3.13), (3.16) and (3.17)).

\section{Auto-B\"acklund transformations}

An auto-B\"acklund transformation of the first order for a  vector equation of the form \eqref{evec} is defined by the formula 
$$
{\bf u}_x = h\, \bsy v_x+f\, \bsy u+g\, \bsy  v,
$$
where ${\bf u}$  and ${\bf v}$ are solutions of \eqref{evec}. The functions $f,g$ and $h$ are (scalar) functions of variables
$$
u_{[0,0]}\stackrel{def}{=}(\bsy  u, \bsy  u),\qquad v_{[i,j]}\stackrel{def}{=}(\bsy v_i, \bsy v_j),\qquad w_{i,j}\stackrel{def}{=}(\bsy  u_i, \, \bsy  v_j), \qquad  i,j\ge 0.
$$

\rem If the auto-B\"acklund transformation depends on an arbitrary parameter $ \mu $, one can construct exact multi-parameter  solutions of equation \eqref{evec} by applying the transformation several times to a trivial solution.

\rem  The existence of a vector  auto-Backlund transformation with an arbitrary parameter is the most easily verifiable evidence for the integrability of a vector equation.

For equations (\ref{e1})--(\ref{e3}) we use the canonical forms  (\ref{e1-aa})--(\ref{e3-aa}) since the auto-B\"acklund  transformations for them look more elegant.

The auto-B\"acklund  transformations for equation (\ref{e1-aa}) with  $\lambda =1$ and with  $\lambda =1/2$ are given by the formulas
\begin{align}\label{e4-aa}
\bsy u_x = \frac{p}{q}\bsy v_x+\frac{p(q\,w_{[0,1]}-p\,v_{[0,1]})}{q^2(a-p\,q+w_{[0,0]})}(\bsy u-\bsy v)-a\,\mu\, \frac{p}{q}(\bsy u-\bsy v),
\end{align}
and
\begin{align}\label{e5-aa}
\bsy u_x = \frac{p}{q}\bsy v_x+\frac{p(q\,w_{[0,1]}-p\,v_{[0,1]}}{q^2(a-p\,q+w_{[0,0]})}(\bsy u-\bsy v)+\frac{\mu\, p^{3/2}(\bsy u-\bsy v)}{q^{1/2}\sqrt{a-p\,q+w_{[0,0]}}}, 
\end{align}
respectively. 
Here   $p=\sqrt{u_{[0,0]}+a}\,\,$, $q=\sqrt{v_{[0,0]}+a}\,\,$ and $\mu$ is an arbitrary  parameter.

The auto-B\"acklund  transformations for equations (\ref{e2-aa})  and (\ref{e3-aa}) have the following form:
\begin{align}\label{e6-aa}
\bsy u_x = \frac{p}{q}\bsy v_x+\frac{\mu\,p\,(p-q)\,(\bsy u-\bsy v)}{a-p\,q+w_{[0,0]}}
\end{align}
and 
\begin{align}\label{e7-aa}
\bsy u_x =\frac{p}{q}\bsy v_x+\frac{\mu\,p\,(\bsy u-\bsy v)}{\sqrt{a-p\,q+w_{[0,0]}}}.
\end{align}

The auto-B\"acklund  transformations  for equation  (\ref{e4}), (\ref{e5}) and (\ref{e7})  have the following form:
\begin{align}\label{B4}
\bsy u_x& =\frac{\sqrt{u_{[0,0]}}}{\sqrt{v_{[0,0]}}}\,\bsy v_x+ \frac{ \sqrt{v_{[0,0]}}\,w_{[0,1]}-\sqrt{u_{[0,0]}}\,v_{[0,1]} }{v_{[0,0]}(w_{[0,0]}- \sqrt{u_{[0,0]}}\, \sqrt{v_{[0,0]}} )}
\big( \bsy u \sqrt{v_{[0,0]}}-\bsy v\sqrt{u_{[0,0]}}\big)+\mu \sqrt{u_{[0,0]}}\,\sqrt{v_{[0,0]}}\,\bsy u,\\[3mm]
\bsy u_x &=\frac{\sqrt{u_{[0,0]}}}{\sqrt{v_{[0,0]}}}\,\bsy v_x+\frac{\sqrt{v_{[0,0]}}\,w_{[0,1]}-\sqrt{u_{[0,0]}}\,v_{[0,1]}}{v_{[0,0]}(w_{[0,0]}-\sqrt{u_{[0,0]}}\, \sqrt{v_{[0,0]}} )}
\big(\bsy u \sqrt{v_{[0,0]}}-\bsy v\sqrt{u_{[0,0]}}\big)+\mu|u_{[0,0]}|^{1/4}|v_{[0,0]}|^{1/4}\bsy u, \label{B5}
\end{align}
and
\begin{align}\label{B6}
\bsy u_x &=\frac{\sqrt{u_{[0,0]}}}{\sqrt{v_{[0,0]}}}\,\bsy v_x+\mu\bsy u\big(\sqrt{u_{[0,0]}}\sqrt{v_{[0,0]}}-w_{[0,0]}\big)^{1/2}
+\mu\frac{ u_{[0,0]}\bsy v- w_{[0,0]}\bsy u }{\big(\sqrt{u_{[0,0]}}\sqrt{v_{[0,0]}}-w_{[0,0]}\big)^{1/2} } \n \\[2.5mm]
&\qquad\qquad\qquad+\frac{\big(\bsy u\sqrt{v_{[0,0]}}-\bsy v\sqrt{u_{[0,0]}}\big)\big(\sqrt{u_{[0,0]}}\,v_{[0,1]} -\sqrt{v_{[0,0]}}\,w_{[0,1]}\big)} {v_{[0,0]}( \sqrt{u_{[0,0]}}\,\sqrt{v_{[0,0]}}-w_{[0,0]})},
\end{align}
respectively. 

\section{Appendix. Geometric properties of equations with $T=0$}

The affine connections that correspond to equation \eqref{e7} and to two equations (\ref{e1-aa}) have zero torsion: $T=0$. We verified that they satisfy the integrability conditions \eqref{Tcon3}-\eqref{jord2}. In the appendix we present explicit formulas for these equations.  

\examp  In the equations (\ref{e1-aa}) we have
$$
A(X,Y) =- \frac{(\bsy u , X)\, Y+ (\bsy u , Y)\, X}{a+u_{[0,0]}},  
$$
where $u_{[0,0]} = (\bsy u, \bsy u).$
One can check that this connection is  the Levi-Civita affine connection of the metric 
$$
g(X,Y)=\frac{(X,Y)}{a+u_{[0, 0]}} -\frac{(\bsy u, X)(\bsy u, Y)}{(a+u_{[0, 0]})^2}.\ 
$$
The tensors $\beta$, $R$ and $\sigma$ can be expressed in terms of $g$ as follows:
$$
\beta(X,Y,Z) = \frac{3\lambda -1}{3}\Big( g(X,Y)\, Z+ g(Y,Z)\, X+g(Z,X)\,Y  \Big),
$$
$$
R(X,Y,Z) = g(X, Z)\,Y- g(X, Y)\, Z \,\, \footnote{This formula means that we are dealing with the space of the constant curvature $k=1$.},
$$
and 
\begin{equation}\label{sigma}
\sigma(X,Y,Z) =\lambda\,g(X,Y) \,Z+\lambda\, g(Z,Y)\, X + (\lambda-1)\, g(X, Z)\, Y.
\end{equation}

It can be easily verified that for any bi-linear form $g$  formula \eqref{sigma} defines a triple Jordan system 
iff  $ \lambda = 1$ or  $\ds \lambda = \frac{1}{2}$ (cf. \eqref{e1-aa}).  Both these triple systems are known to be simple \cite{Zel}.

\examp  An elegant description \cite{SviSok96} of all geometric objects for equation (3.8) can be done in terms of the simple triple Jordan system (cf. \eqref{sigma})
$$
S(X,Y,Z) \stackrel{def}{=} (X, Y) \, Z + (Z, Y) \, X - (X, Z) \, Y.  
$$
We have  
$$
A(X, Y) = S(X, F, Y), \qquad \mbox{where} \quad  F \stackrel{def}{=} -\frac{\bsy u}{u_{[0,0]}},
$$
$$
\sigma(X,Y,Z) =  -\frac{1}{2}\, S(X,\, S(F, Y, F),\, Z), $$$$
R(X,Y,Z) = \sigma(X,Z,Y) - \sigma(X,Y,Z), \qquad  \beta(X,X,X) = \sigma(X,X,X).
$$
The tensor $\beta(X,Y,Z)$ can be obtained from $\beta(X,X,X)$  by the symmetrization.

\section{Appendix. Integrable triangular systems}\label{app2} 
Since $T=0$ for triangular systems of the form \eqref{eq13}, we may use the intgerability conditions \eqref{Tcon3} - \eqref{jord2} for the classification of triangular systems.

\lemm Using a transformation of the form \eqref{point}, we can reduce the coefficient $a_5$ in \eqref{eq13} to 
\begin{itemize}
\item {\bf Case a}: $\qquad  a_5=0$;
\item {\bf Case b}:  $\ds \qquad  a_5(x)=\frac{3}{x}.$
\end{itemize}
In the {\bf Case a} the conditions \eqref{Tcon3} - \eqref{jord2} are equivalent to $a_6=a_7=a_8=a_9=0$ and we arrive at the systems
$$
\bsy u_t=\bsy u_3-3\bsy u_2\frac{u_{[0,1]}}{u_{[0,0]}}+\bsy u_1\left((9-c_1)\frac{u_{[0,1]}^2}{u_{[0,0]}^2}-3\frac{u_{[0,2]}}{u_{[0,0]}}+(c_1-3)\frac{u_{[1,1]}}{u_{[0,0]}}\right), \qquad c_1 = 3,\, \frac{3}{2}. 
$$
In the {\bf Case b} we may use transformations \eqref{point} to vanish $a_6$. Transformations \eqref{point} with 
\begin{equation}\label{ff}
f(x)=k_1 x^{k_2},
\end{equation}
where $k_i$ are arbitrary constants, preserve the normalization $a_6=0.$ From conditions \eqref{Tcon3} - \eqref{jord2} it follows that 
$$\ds {\bf Case} \,\,  {\bf b_1}: \quad a_7(x) = -\frac{3}{x^2} \qquad \mbox{or} \quad {\bf Case} \,\, {\bf b_2}: \quad a_7(x) = -\frac{c_1+3}{x^2}.$$

In the Case $b_1$ conditions \eqref{Tcon3} - \eqref{jord2} imply $\ds c_1=\frac{3}{2}$ and $\ds a_8(x)=-\frac{1}{x^3}$ and we obtain the equation 
$$
\bsy u_t=\bsy u_3-3\bsy u_2\frac{u_{[0,1]}}{u_{[0,0]}}+\bsy u_1\left(\frac{15}{2}\frac{u_{[0,1]}^2}{u_{[0,0]}^2}-3\frac{u_{[0,2]}}{u_{[0,0]}}-\frac{3}{2}\frac{u_{[1,1]}}{u_{[0,0]}}\right) + 
\bsy u \left(3\frac{u_{[1,2]}}{u_{[0,0]}}-3\frac{u_{[0,1]} u_{[1,1]}}{u_{[0,0]}^2}- \frac{u_{[0,1]}^3}{u_{[0,0]}^3}\right).
$$ 
This equation is invariant with respect to the group of transformations \eqref{point}, \eqref{ff}.

In the Case $b_2$ we get $\ds a_8 = \frac{k}{x^3},$ where $k$ is a constant. By a transformation \eqref{point}, \eqref{ff} we can bring $k$ to zero. As a result we obtain 
$$
\begin{array}{c}
\ds \bsy u_t=\bsy u_3-3\bsy u_2\frac{u_{[0,1]}}{u_{[0,0]}}+\bsy u_1\left((9-c_1)\frac{u_{[0,1]}^2}{u_{[0,0]}^2}-3\frac{u_{[0,2]}}{u_{[0,0]}}+(c_1-3)\frac{u_{[1,1]}}{u_{[0,0]}}\right) + \\[4mm] 
\ds \bsy u \left(3\frac{u_{[1,2]}}{u_{[0,0]}}-(c_1+3)\frac{u_{[0,1]} u_{[1,1]}}{u_{[0,0]}^2} \right)
 \qquad c_1 = 3,\, \frac{3}{2}. 
\end{array}
$$
Both of these equations admit a {\bf total} separation of variables in the spherical coordinates: the equation with $c_1=3$ is converted to 
$$
\bsy v_t =\bsy v_{xxx}+ 3 \bsy v_x v_{[1, 1]}+3 \bsy v v_{[1, 2]},   \qquad R_t = R_{xxx}- 3\frac{R_x R_{xx}}{R}
$$
while the equation with $\ds c_1=\frac{3}{2}$ turns into
$$
\bsy v_t =\bsy v_{xxx}+ \frac{3}{2} \bsy v_x v_{[1, 1]}+3 \bsy v v_{[1, 2]},   \qquad R_t = R_{xxx}- 3\frac{R_x R_{xx}}{R} + \frac{3}{2}\frac{R_x^3}{R^2}.
$$
In both cases the scalar equation for $R$ is point equivalent to the integrable equation $\tilde R_t = \tilde R_{xxx} + \tilde R_x^3$.
 
The equation from Case $b_1$ and  the equations from Case {\bf a} admit a {\bf partial} separation of variables in the spherical coordinates.

\end{document}